\documentclass[aps,pra,onecolumn,groupedaddress,showpacs]{revtex4}

\usepackage{hyperref}
\usepackage{graphics}
\usepackage{amssymb,amsmath}
\draft
\usepackage{hyperref}
\usepackage{graphicx}
\newcommand{\Om}{\Omega}

\newcommand{\om}{\omega}
\newcommand{\al}{\alpha}

\newcommand{\ve}{\varepsilon}
\newcommand{\pa}{\partial}

\newcommand{\vkp}{\kappa}

\begin{document}

\title{Vector pulsing solitons in Kerr media }

\author{G. T. Adamashvili, N. T. Adamashvili, M. D. Peikrishvili, R. R. Koplatadze}

\vskip+0.2 cm
\affiliation{Technical University of Georgia, Kostava str.77, Tbilisi, 0179, Georgia. \\ email: $guram_{-}adamashvili@ymail.com.$ }

\begin{abstract}
A theory of optical nonresonance vector pulsing solitons in a Kerr media is considered. By using the perturbative reduction method the wave equation is transformed to the coupled nonlinear Schrödinger equations.
The profile of the optical nonresonance vector pulsing soliton with the difference and sum of the frequencies is presented.
Explicit analytical expressions for the optical two-component vector pulsing soliton with phase modulation are obtained.
It is shown that the two-component pulsing soliton in this special case can be transformed to the scalar pulsing soliton,
and these waves have different profiles.
\end{abstract}

\pacs{42.65.Tg}

\maketitle

\centerline{\textbf{1. Introduction}}

\vskip+0.5 cm

The propagation of optical waves in a medium is accompanied by different changes in their profile. The effects
changing the wave profile are dispersion, dissipation and nonlinearity. These mechanisms act separately or in different
combinations. Of special interest are such wave motions for which the mechanisms distorting the profile and induced by
different effects exactly compensate each other. Under these conditions, nonlinear waves of stationary profile such as solitons
or their different modifications are formed. The propagation of nonlinear waves of an invariable profile displays its
own specific properties. In the theory of nonlinear waves they play as fundamental a role as harmonic oscillations do
in the linear wave theory. The nonlinear waves of an invariable profile are one of the most important demonstrations of
nonlinearity in optical systems.

The conditions for the existence of nonlinear waves are different. The determination of
the conditions of the existence of optical nonlinear waves of a stationary profile and the study of their features in different
physical situations are among the principal problems of the nonlinear wave theory. Depending on the character of the
nonlinearity, the nonresonance and resonance mechanism of the existence of nonlinear waves is realized. In the first case
of nonresonant nonlinearity, which is expressed by means of the quadratic or cubic nonlinear susceptibilities, its competition
with the dispersion leads to the existence of nonresonance optical solitons and pulsing solitons (breathers) \cite{Sauter::96, AdamashviliMaradudin:PhysRevE:97, AdamashviliKaup:PhysRevE:04}. The optical resonant nonlinear solitary waves can be excited
with the help of self-induced transparency, i.e., from a coherent nonlinear resonance interaction of an optical wave
with impurity atoms or semiconductor quantum dots in solids \cite{Allen::75, Adamashvili:OptLett:06}.

Nonlinear solitary waves can be considered by a single nonlinear Schrödinger (NLS) equation for the optical one-component
(scalar) field. Such one-component resonance and nonresonance nonlinear waves form when an optical one-component pulse propagates inside a medium while maintaining its state \cite{Adamashvili:PhysRevE:04, AdamashviliKaup:PhysRevE:04}. When this is not the case, the interaction between two field components at different polarizations or different frequencies (but possibly same polarization)
has to be considered. One then has to simultaneously solve a system of coupled NLS equations. A profile preserving solution of the coupled NLS equations is a vector pulse (soliton or pulsing soliton) because of its two-component configuration.

The properties of optical nonresonance vector pulsing solitons in a Kerr medium are governed by two coupled NLS equations
that describe the connection between two different guided modes propagating in multi-mode optical waveguides
(fibers) \cite{Crosignani::81} or the coupling between two optical wave components of two distinct carrier frequencies propagating
inside a single-mode waveguide \cite{Agrawal::01}. In addition, in a single-mode waveguide, a single pulse also can form a vector soliton if the birefringence effects lead to a connection between its two differently polarized wave components \cite{Menyuk::89}.

It is of great importance to find double periodic vector soliton (vector pulsing soliton) solutions of optical nonlinear
equations to provide more important information for understanding phenomena arising in different scientific fields and
applications. The pulsing solitons have various interesting features that are similar to those of solitons, but unlike them,
pulsing solitons can be created with relatively low input pulse energy. Therefore, pulsing solitons are easier to excite
than solitons and, in addition, in some physical phenomena, pulsing solitons are more stable nonlinear waves and thus
have wider potential applications in comparison to solitons (see, for example, Ref. \cite{Chen:Phys. Rev. E :04}).

The theory of two-component nonresonance vector pulsing solitons with the difference and sum of the frequencies in
a Kerr medium will be different from and more complex than the nonresonance one-component solitons and pulsing
solitons \cite{Sauter::96, AdamashviliMaradudin:PhysRevE:97, AdamashviliKaup:PhysRevE:04} and two-component vector solitons
\cite{Crosignani::81,Agrawal::01, Menyuk::89}, and a separate study will be needed.

The goal of the present work is the following: we consider the conditions of realization of the nonresonance two-component
optical vector pulsing soliton with the difference and sum of the frequencies with a phase modulation in a Kerr medium. We explicitly determine analytic expressions for the parameters and the profile of the optical nonresonance vector pulsing soliton.

\vskip+0.5 cm

\centerline{ \textbf{2. Basic equations}}

\vskip+0.5 cm

We study the propagation of optical nonresonance two-component vector pulsing solitons in isotropic, cubic nonlinear
and second order dispersive media for linearly polarized waves with width $T$ and frequency $\omega\gg T^{-1}$ with the
strength of the electric field $\vec{E}(z,t)=\vec {e}\;E(z,t)$  propagating along the positive $\emph{z}$-axis, where $\vec {e}$ is the unit vector of polarization directed along the $\emph{x}$-axis.

Not concretizing the physical nature of the dispersive process, we describe the dependence of the dielectric function $\vkp$  by a two variables: wave vector $\vec{k}$ and frequency $\omega $ of the wave (spatial and/or temporal dispersion). We note that in optical phenomena we usually consider only temporal dispersion, but in some special physical situations, spatial dispersion can be effective too (see, for instance, Refs. \cite{AdamashviliMaradudin:PhysRevE:97, AdamashviliKaup:PhysRevE:04, Vinogradova::90} and references therein).

The nonlinear wave equation has the following form \cite{Landau:Electrodynamics :84, Vinogradova::90}:
\begin{widetext}
\begin{equation}\label{eq.1}
- C^2 \frac{\pa^{2} E}{\pa {z}^2} +\frac{\pa^{2} }{\pa t^2}\int
\kappa (z_1,t_1) E(z-z_1,t-t_1)dz_1 dt_1
+4\pi\frac{\pa^{2} P}{\pa t^2}=0,
\end{equation}
where
\begin{equation}\label{eq.2}
P=\int\rho_{xxxx}({z}_1,{z}_2,{z}_3,t_1,t_2,t_3)E(z-z_1,t-t_1)\times $$$$
E(z-z_1-z_2,t-t_1-t_2)E(z-z_1-z_2-z_3,t-t_1-t_2-t_3) dz_1 dz_2
dz_3 dt_1 dt_2 dt_3
\end{equation}
\end{widetext}
is the x-component of the non-resonant nonlinear polarization of the  third order,
 $\rho_{xxxx}$ is the component of the tensor of the cubic susceptibility, $C$ is  the light velocity in vacuum, $\kappa (z,t)=1+ 4\pi \chi_{xx}(z,t)$,
$\chi_{xx}$ is the component of the first-order susceptibility tensor.

We can simplify Eq.\eqref{eq.1} with the use of the slowly changing profiles method. In order to do this, we represent
the functions $E$  as
\begin{equation}\label{eq.3}
E = \sum_{l=\pm 1}\hat{E}_l Z_l,\;\;\;
\end{equation}
where $\hat{E}_l$  is the slowly varying complex amplitude of the optical electric field, $Z_l=e^{il(k{z}- \om t)}$. To guarantee that $E(z,t)$ is a real function, we suppose that $\hat{E}_{l}=\hat{E}_{-l}^{\ast}$.
In comparison with the carrier wave parts, the complex envelope functions $\hat{E}_{l}$ vary slowly in space and time, i.e.,
\begin{equation}\label{eq.14a}\nonumber\\
 \left|\frac{\partial \hat{E}_{l}}{\partial t}\right|\ll\omega
|\hat{E}_{l}|,\;\;\;\left|\frac{\partial \hat{E}_{l}}{\partial z
}\right|\ll k|\hat{E}_{l}|.
\end{equation}

Substituting the equation \eqref{eq.3} into the equation \eqref{eq.2}, we obtain
\begin{equation}\label{eq.3a}
P=\sum_{l,l',l''} Z_{l}\tilde{\rho}_{l,l',l''}
 {\hat{E}_{l-l'-l''}}{\hat{E}_{l'}}{\hat{E}_{l''}},
\end{equation}
where
$$
\tilde{\rho}_{l,l',l''}=\int\rho_{xxxx}({z}_1,{z}_2,{z}_3,t_1,t_2,t_3)
 e^{-il(kz_1- \om t_1)}
e^{-i(l'+l'')[k z_2- \om t_2]} e^{-il''[k z_3- \om t_3]}dz_1 dz_2
dz_3 dt_1 dt_2 dt_3.
$$

Substituting the equations \eqref{eq.3} and  \eqref{eq.3a} into the wave equation \eqref{eq.1}, we obtain  the dispersion law for propagating pulse
\begin{equation}\label{er19}
C^{2}k^{2}= {\om}^2{\vkp}(k,\om)
\end{equation}
and a nonlinear wave equation in the form:
\begin{equation}\label{rty2}
\sum_{l=\pm1}Z_{l}\{[ig_{1} \frac{\partial
\hat{E}_{l}}{\partial z}+g_{2} \frac{\partial^{2}\hat{E}_{l}}{\partial z^2}+ig_{3} \frac{{\pa}{\hat{E}_l}}{\pa t}  +g_{4}\frac{{{\pa}^2}{\hat{E}_l}}{\pa t^2}    -g_{5}\frac{{{\pa}^2}{\hat{E}_l}}{{\pa {z}}{\pa
   t}}]-\sum_{l'}\sum_{l''}  K_{l,l',l''}{\hat{E}_{l-l'-l''}}{\hat{E}_{l'}}
{\hat{E}_{l''}}\}=0,
\end{equation}
where
\begin{equation}\label{r2}\nonumber\\
g_{1}={\om}^2 a -2lk C^{2},\;\;\;\;\;\;\;\;
\;\;\;g_{2}={\om}^2 c -C^{2},\;\;\;\;\;\;\;\;g_{3}=-2 l{\om} \vkp - {\om}^{2} b,\;\;\;\;\;\;\;\;
$$$$
g_{4}={\vkp}+2l{\om}b+{\om}^2 d ,\;\;\;g_5=2l{\om} a +{\om}^{2} t,
\;\;\;\;\;\;\;\;
K_{l,l',l''}=4\pi \omega^{2}\tilde{\rho}_{l,l',l''},
\end{equation}
$$
       a=\frac{\pa {{\vkp}}}{\pa(lk)},\;\;\;
  b=  \frac{\pa {{\vkp}}}{\pa(l\om)},\;\;\;
  c= \frac{1}{2}\frac{\pa^2 {{\vkp}}}{\pa(lk)^2},\;\;\;
    d=  \frac{1}{2}\frac{\pa^2 {{\vkp}}}{\pa(l\om)^2},\;\;\;
     t= \frac{\pa^2 {{\vkp}}}{\pa(lk)\pa(l\om)}.
$$
$$
\vkp(k,\om)=\int \vkp(z_1,t_1)e^{-il(kz_1-\om t_1)} dz_1 dt_1.
$$
The function $\vkp (k,\om)$ is in general complex, but we consider
only the most important particular case when a wave propagated without damping in a non-absorbing (transparent) homogeneous medium. In this case real part of $\vkp $ is an even function of the frequency and wave number and  imaginary part of this function equal to zero \cite{Vinogradova::90, Landau:Electrodynamics :84}.

\vskip+0.5 cm
\centerline{ \textbf{3. Nonresonance vector pulsing soliton}}

\vskip+0.5 cm

To further analyze of these equations we make use of the multiple scale perturbative reduction method \cite{Taniuti::1973}, in the limit that $ \hat{E}_{l}(z,t)$ is of order ${\cal O}(\epsilon)$, where $\varepsilon$ is a small parameter. In this case $ \hat{E}_{l}(z,t)$ can be represented as:
\begin{equation}\label{rty3}
\hat{E}_{l}(z,t)=\sum_{\alpha=1}^{\infty}\sum_{n=-\infty}^{+\infty}\varepsilon^\al
Y_{l,n} f_{l,n}^ {(\alpha)}(\zeta,\tau),
\end{equation}
where
$$
Y_{l,n}=e^{in(Q_{l,n}z-\Omega_{l,n}
t)},\;\;\;\zeta_{l,n}=\varepsilon Q_{l,n}(z- v_{l,n}
t),\;\;\;\tau=\varepsilon^2 t,\;\;\;
{v}_{l,n}=\frac{d\Omega_{l,n}}{dQ_{l,n}}.
$$
Such a representation allows
us to separate from $ \hat{E}_{l}$ the still more slowly changing
quantities $ f_{l,n}^{(\alpha )}$. Consequently, it is assumed that
the quantities $\Omega_{l,n}$, $Q_{l,n}$, and $f_{l,n}^{(\alpha)}$ satisfy the
inequalities for any $l$ and $n$:
\begin{equation}\label{rtyp}\nonumber\\
\omega\gg \Omega,\;\;k\gg Q,\;\;\;
\end{equation}
$$
\left|\frac{\partial
f_{l,n}^{(\alpha )}}{
\partial t}\right|\ll \Omega \left|f_{l,n}^{(\alpha )}\right|,\;\;\left|\frac{\partial
f_{l,n}^{(\alpha )}}{\partial z }\right|\ll Q\left|f_{l,n}^{(\alpha )}\right|.
$$
We have to note that the quantities  $Q$, $\Omega$ and $v$ depends from
$l$ and $n$, but for simplicity, we omit these indexes in equations where this will not bring about mess.

Substituting Eq.\eqref{rty3} into Eq.\eqref{rty2} we obtain
\begin{widetext}
\begin{equation}\label{uq9}
\sum_{l=\pm1}\sum_{\alpha=1}^{\infty}\sum_{n=-\infty}^{+\infty}\varepsilon^\al
Y_{l,n} Z_l\{W_{l,n}+ i \varepsilon J_{l,n} \frac{\partial }{\partial \zeta}
+ i\varepsilon^2 h_{l,n}
\frac{\partial }{\partial \tau}
+\varepsilon^{2} H_{l,n} \frac{\partial^2 }{\partial\zeta^2}\}f_{l,n}^{(\alpha)}$$$$=\ve^3 \sum_{{l=\pm 1}}Z_l  R_{l} [( | f_{l,l}^ {(1)}|^{2}   +
 2  | f_{l,-l}^ {(1)}|^{2}  ) f_{l,l}^ {(1)}  Y_{l,l}
+(  |f_{l,-l}^ {(1)}|^{2}  + 2  | f_{l,l}^ {(1)}|^{2} )f_{l,-l}^ {(1)} Y_{l,-l}],
\end{equation}
where
\begin{equation}\label{uq20a}
W_{l,n}= g_{3}n\Omega -g_{1} nQ - g_{2} Q^{2} -g_{4} {\Omega}^{2} -g_{5} Q \Omega,
$$$$
J_{l,n} =g_{3}  v -g_{1} - 2g_{2} nQ - 2 n g_{4}  \Omega  v
-g_{5}n(Q v +\Omega),
$$$$
h_{l,n}=g_{3} - 2 g_{4}n\Omega
-g_{5} nQ,
$$$$
H_{l,n}=Q^{2}(g_{2} +g_{4}v^{2}+g_{5} v ),
$$$$
R_{l}=4\pi \omega^{2} (\tilde{\rho}_{l,l,l}
+\tilde{\rho}_{l,l,-l} +\tilde{\rho}_{l,-l,l}).
\end{equation}
\end{widetext}

To determine the values of $f_{l,n}^{(\alpha)}$, we equate to zero the
various terms corresponding to the same powers of $\ve$. As a
result, we obtain a chain of equations. Starting with first order in $\ve$, we have
\begin{equation}\label{uq20}
\sum_{l,n=\pm1} Z_{l} Y_{l,n}{W}_{l,n}f_{l,n}^{(1)}=0.
\end{equation}
Consequently, according to Eq.\eqref{uq20}, only the following components of $f_{l,n}^{(1)}$ can differ from zero: $f_{+ 1,\pm 1}^{(1)}$ or $f_{+ 1,\mp 1}^{(1)}$. From the condition $\hat{E}_{l}=\hat{E}_{-l}^{\ast}$ follows, that $ {f^{*}}_{+1,\pm 1}^ {(1)}= f_{-1,\mp 1}^ {(1)}$.
The relations between the parameters $\Omega $ and $Q$ is determined from Eq.\eqref{uq20} and has the form
\begin{equation}\label{e16}
g_{3}n\Omega -g_{1} nQ - g_{2} Q^{2} -g_{4} {\Omega}^{2} -g_{5} Q \Omega=0.
\end{equation}

Substituting  Eq.\eqref{e16} into Eqs. \eqref{uq9} and \eqref{uq20a}, to second order in $\ve$, we easily see that the following relation holds $J_{+1,\pm 1}=J_{+ 1,\mp 1}=0$.

From the Eq.\eqref{uq9}, to third order in $\ve $, we obtain the following nonlinear equation
\begin{equation}\label{eq1w}
i h_{+1,\pm 1} \frac{\partial f_{+1,\pm 1 }^{(1)}}{\partial \tau}
+ H_{+1,\pm 1} \frac{\partial^2 f_{+1,\pm 1}^{(1)}}{\partial\zeta^2}- R_{+1}  | f_{+1,\pm 1}^ {(1)}|^{2} f_{+1,\pm 1}^ {(1)}  - 2 R_{+1}  | f_{+1,\mp 1}^ {(1)}|^{2}   f_{+1,\pm 1}^ {(1)}=0.
\end{equation}

From Eq.\eqref{eq1w}, we finally obtain two coupled NLS equations  for functions $U={\ve}f_{+1,+1}^{(1)}$ and $V={\ve}f_{+1,-1}^{(1)}$ that describe the coupling between two components of the pulse
\begin{equation}\label{eq2w}
 i  (\frac{\partial U}{\partial t}+ v_{+1} \frac{\partial U}{\partial z})
+  p_{+1} \frac{\partial^{2} U
}{\partial z^{2}} + q_{+1} (|U|^{2} +2 |V|^{2}) U=0,
$$$$
 i  (\frac{\partial V}{\partial t}+ v_{-1} \frac{\partial V}{\partial z})
+  p_{-1} \frac{\partial^{2} V
}{\partial z^{2}} + q_{-1} (|V|^{2}+2|U|^{2}) V=0,
\end{equation}
where
\begin{equation}\label{eq3w}
v_{\pm 1}=v_{+1,\pm 1},\;\;\;\;\;\;\;\;\;\;\;\;\;
p_{\pm 1}= \frac{H_{+1,\pm 1}}{h_{+1,\pm 1} Q_{+1,\pm 1}^{2}},\;\;\;\;\;\;\;\;\;\;\;\;\;\;\;
q_{\pm 1}= -\frac{R_{+1}}{h_{+1,\pm 1}},
$$$$
\Omega_{+1}=\Omega_{l=\pm1,n=\pm1},\;\;\;\;\;\;\;\;\;\Omega_{-1}=\Omega_{l=\pm1,n=\mp1},
$$$$
Q_{+1}=Q_{l=\pm1,n=\pm1},\;\;\;\;\;\;\;\;\;Q_{-1}=Q_{l=\pm1,n=\mp1}.
\end{equation}

The nonlinear equations \eqref{eq2w}  describes the slowly varying envelope functions $U$ and $V$,
where $U$ describes the envelope wave of the frequency $\om+\Om_{+1}$ and $V$ describes the wave with frequency $\om-\Om_{-1}$. The nonlinear coupling between the two waves is governed by the terms $|V|^{2}U$ and $|U|^{2}V$. We must consider interaction of these field components at different frequencies and the same polarization and solve simultaneously  a set of coupled NLS equations \eqref{eq2w}. A shape-preserving solution of the equations \eqref{eq2w} is a vector pulse because of its two-component structure.

We have to note, that the coupled NLS equations  arise in different fields of physics.  At this,  depending on the physical situations,  there are different relations between coefficients of these equations and can be different ways for their solutions. The simplest way to ensure the steady-state property is to require the field envelope functions to depend on the time and space coordinate only through the coordinate $ \xi = t- \frac{z}{V_{0}}$, where $V_{0}$ is the constant vector pulse velocity. We are looking for the steady-state solutions for the strength of the electrical  field of the pulse and therefore solution of the Eqs.\eqref{eq2w} for complex amplitudes we will search in the form \cite{Adamashvili:Eur.Phys.J.D.:12}:
\begin{equation}\label{eq12}
U(z,t)= \frac{A_{+1}}{bT}e^{i\phi_{+1}} sech \frac{\xi}{T} ,\;\;\;\;\;\;\;\;\;\;V(z,t)=\frac{A_{-1}}{bT}e^{i\phi_{-1}} sech \frac{\xi}{T},
\end{equation}
where $b^{2}=  \frac{V_{0}^{2}\;q_{+1}}{2p_{+1}}(A_{+1}^{2} +2 A_{-1}^{2}),\;\;$$\phi_{\pm 1}=k_{\pm 1} z- \omega_{\pm 1} t$ are the phase functions, $A_{\pm 1},\;$$\;k_{\pm 1},\;$ and $\omega_{\pm 1}$ are all real constants. Derivatives of the phase $\phi_{\pm 1}$ are assumed to be small, i.e. the functions $ e^{i\phi_{\pm 1}}$ are  slow in comparison with oscillations of the pulse and consequently, the inequalities
\begin{equation}\label{eq12a}\nonumber\\
k_{\pm 1}=\frac{V_{0}-v_{\pm 1}}{2p_{\pm 1}}<<Q_{\pm 1},\;\;\;\;\omega_{\pm 1}<<{\Omega}_{\pm 1}
\end{equation}
are satisfied. The relations between quantities $A_{+1},A_{-1}$ and $\omega_{+1},\omega_{-1}$ have the following form:
\begin{equation}\label{rt16}
A_{-1}^{2}=\frac{p_{-1}q_{+1}- 2 p_{+1}q_{-1}}{p_{+1}q_{-1}-2p_{-1}q_{+1}}A_{+1}^{2},$$$$
\;\;\;\;\;\;\;\;\;\;\;\;\;\;\;\;\;
\omega_{+1}=\frac{p_{+1}}{p_{-1}}\omega_{-1}+\frac{V^{2}_{0}(p_{-1}^{2}-p_{+1}^{2})+v_{-1}^{2}p_{+1}^{2}-v_{+1}^{2}p_{-1}^{2}
}{4p_{+1}p_{-1}^{2}}.
\end{equation}

Substituting the solutions for the functions $U$  and $V$  Eq.\eqref{eq12}  of the coupled NLS equations \eqref{eq1w}  into Eqs.\eqref{eq.3} and \eqref{rty3}, we obtain for the \emph{x}-component of the electric field strength $E(z,t)$ the two-component vector breather solution of the Maxwell equation \eqref{eq.1} in the form:
\begin{equation}\label{eq17}
E(z,t)=
\frac{2}{b T}sech(\frac{t-\frac{z}{V_{0}}}{T})\sum_{j=\pm 1}A_{J} [cos (\omega+ J \Omega_{J}+\omega_{J})t - (k + JQ_{J}+k_{J})z ],
\end{equation}
where
\begin{equation}\label{er17}
T^{-2}=V_{0}^{2}\frac{v_{+1}k_{+1}+k_{+1}^{2}p_{+1}-\omega_{+1}}{p_{+1}}.
\end{equation}

\begin{figure}[htbp]
\includegraphics[width=0.99\textwidth]{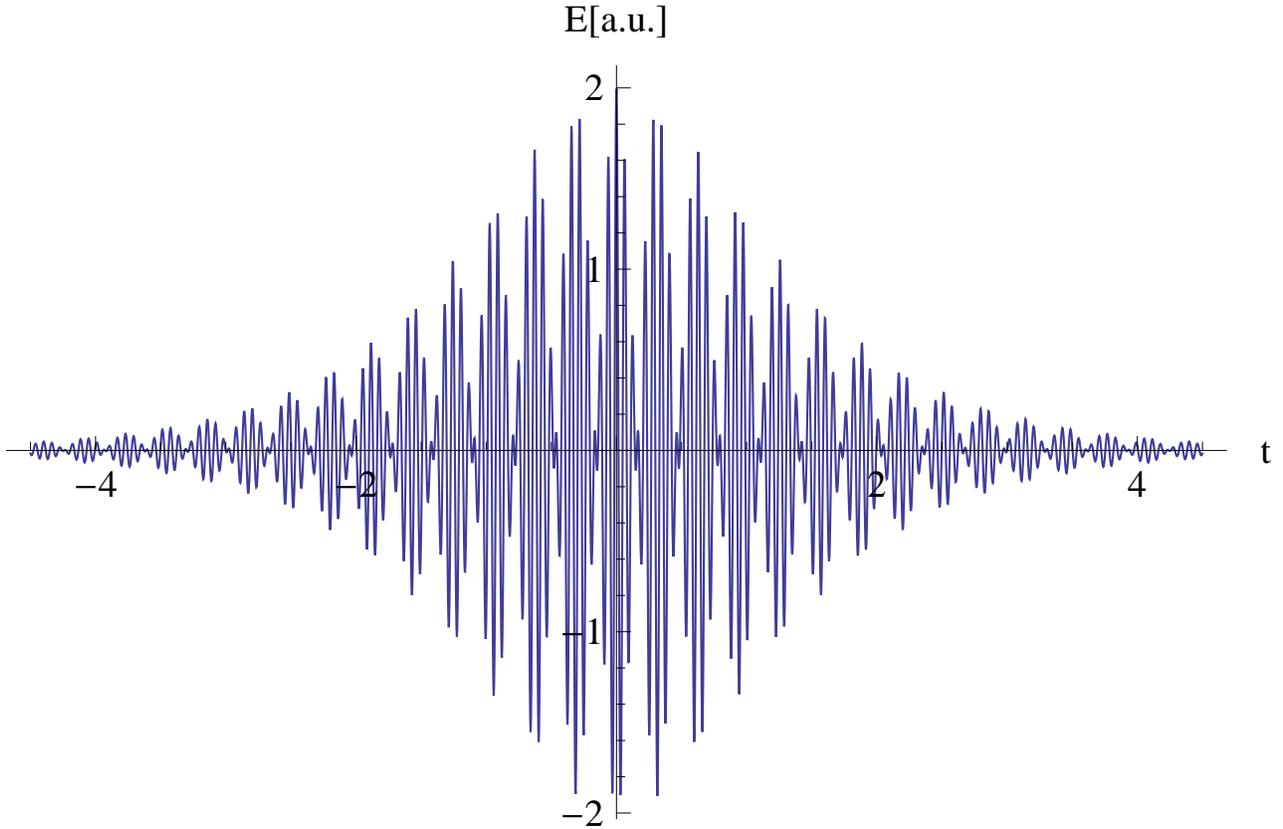}
\caption{The $x$-component of the strength of the electrical field $E(0,t)$ of the two-component vector pulsing soliton is shown for a fixed value of $z$. The nonlinear pulse oscillates with the difference $\om -\Omega_{-1}$ and sum $\om +\Omega_{+1}$
of the frequencies along the $t$-axis.}
\label{fig1}
\end{figure}

\begin{figure}[htbp]
\includegraphics[width=0.99\textwidth]{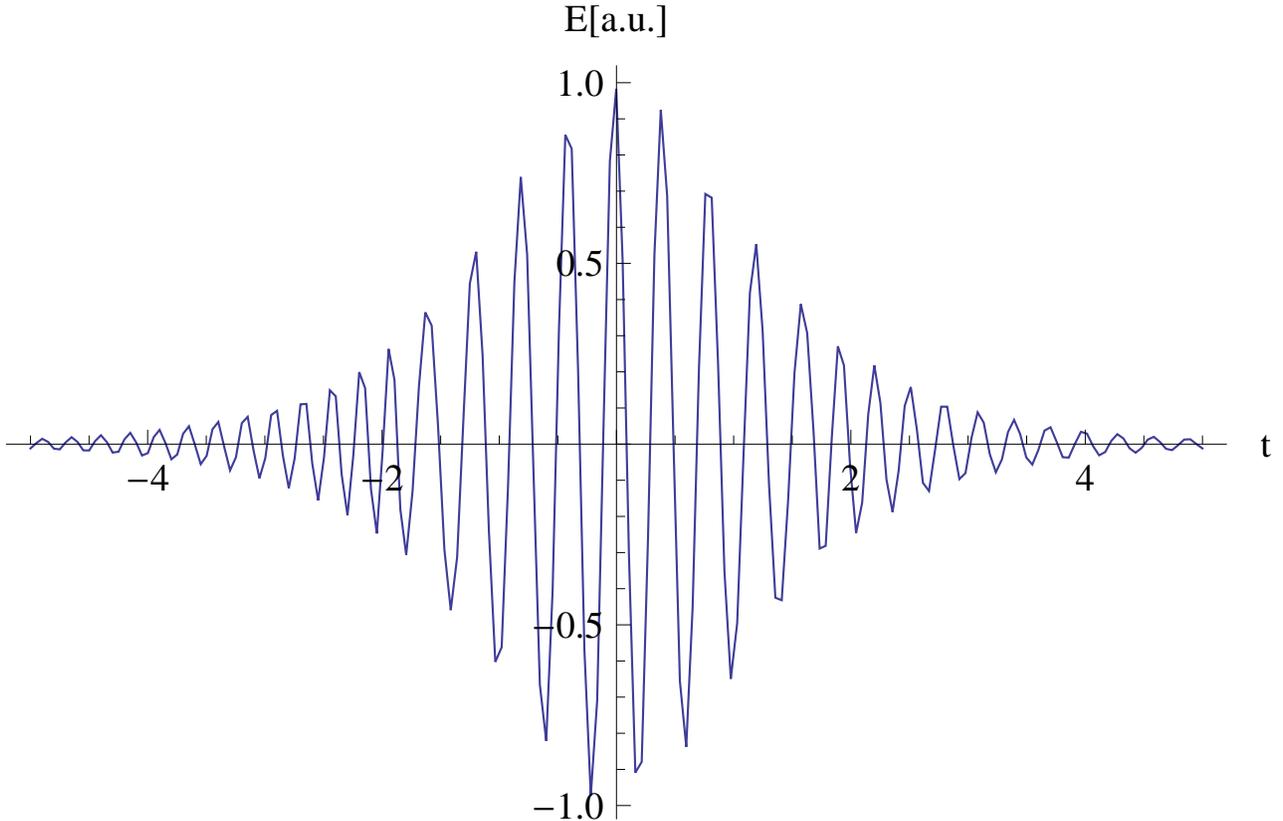}
\caption{The $x$-component of the strength of the electrical field $E(0,t)$ of the one-component scalar pulsing soliton is shown for a fixed value of $z$.}
\label{fig1}
\end{figure}

\vskip+0.5 cm

\centerline{ \textbf{4. Conclusion}}

\vskip+0.5 cm

We demonstrated that in cubic nonlinear and second order (spatially and/or temporally) dispersive media, an optical
nonresonance two-component vector pulsing soliton can arise. The explicit expressions for the parameters and profile
of the optical nonresonance vector pulsing soliton are given by Eqs.\eqref{eq3w}, \eqref{rt16}, \eqref{eq17} and \eqref{er17}. The dispersion law and the connection between the quantities $\Omega_{\pm 1}$ and $Q_{\pm 1}$ are given by Eqs.\eqref{er19} and \eqref{e16}, respectively.

In Eq. \eqref{eq17}  the functions  $ cos[ (\om +\Omega_{+1}+\omega_{+1}) t -(k+Q_{+1}+k_{+1})z]$ and $cos[(\om -\Omega_{-1}+\omega_{-1})t- (k-Q_{-1}+k_{-1})z ]$ indicates the formation of double periodic beats with coordinate and time relative to the frequency and wave number of the carrier wave ($\omega$, $k$), with  characteristic parameters ($\om +\Omega_{+1}$, $k+Q_{+1}$) and ($\om -\Omega_{-1}$, $k-Q_{-1}$), respectively. Eq.\eqref{eq17} is exact regular time and space double periodic solution of nonlinear equation \eqref{eq.1}  which, like a one-component soliton and breather loses no energy during propagation through the medium.

A vector pulsing soliton is an absolutely different nonlinear wave in comparison with nonresonance one-component
solitons and breathers Refs.\cite{Sauter::96, AdamashviliKaup:PhysRevE:04} which have been investigated up to now. The vector pulsing soliton is a complex nonlinear wave which consists of two coupled pulsing solitons with different frequencies of oscillation
and the same polarizations (along the $x$-axis), which in the process of propagation exchange the energy between
each other.

A plot of the two-component vector pulsing soliton Eq.\eqref{eq17} is shown in Fig.1 for a fixed value of the $z$ coordinate.
We assume that the quantities $\omega/\Omega_{\pm 1}$ are of the order $10^{3}$. In the optical region of the spectrum, $\omega$ is of the order  $10^{15}$  $s^{-1}$ and with a typical numerical values for the pulse width in glass  $T=100$  ps \cite{Sauter::96}, the condition $T \;\Omega_{\pm 1}>>1$ is fulfilled.

The one-component pulsing soliton is a special case of the vector pulsing soliton Eq.\eqref{eq17}. The shape of the one-component
pulsing soliton with the same values of the parameters as for the vector pulsing soliton is presented in Fig.2. It is obvious that the profile of the two-component vector pulsing soliton (Fig.1) differs from the profile of the one-component
pulsing soliton (Fig.2).

The results of this theoretical study of the non-resonance vector breathers, together with
those obtained in Refs.\cite{Sauter::96, AdamashviliKaup:PhysRevE:04} for the one-component solitons and breathers,  provides a more complete physical description  the propagation of the non-resonance nonlinear waves in Kerr media.

It should be noted that the constructed theory is quite general and can be transformed for second order (spatially and/or temporally) dispersive and  noncentrosymmetric crystals with quadratic susceptibility.

\end{document}